\documentclass[aps,prd,reprint,twocolumn,superscriptaddress,longbibliography,nofootinbib,floatfix,showpacs]{revtex4-1}
\usepackage{epsfig}
\usepackage{amsmath,amssymb,amsfonts}
\usepackage{hyperref}
\usepackage{mathrsfs}
\usepackage{bbm}
\usepackage{slashed}
\usepackage{graphicx}
\usepackage{verbatim}
\usepackage[usenames]{color}

\newcommand{\be}{\begin{equation}}
\newcommand{\ee}{\end{equation}}
\newcommand{\bi}{\begin{itemize}}
\newcommand{\ei}{\end{itemize}}
\newcommand{\bea}{\begin{eqnarray}}
\newcommand{\eea}{\end{eqnarray}}

\newcommand{\bracket}[2]{\bra{#1}\,#2\rangle} % Dirac inner product
\newcommand{\bra}[1]{\langle\,#1\,|}          % Dirac bra
\newcommand{\ket}[1]{|\,#1\,\rangle}          % Dirac Ket

\newcommand{\ud}{\mathrm{d}}

\newcommand{\LCm}{{\scriptscriptstyle -}} %LC supersripts
\newcommand{\LCp}{{\scriptscriptstyle +}}
\newcommand{\LCpm}{{\scriptscriptstyle \pm}}
\newcommand{\LCmp}{{\scriptscriptstyle \mp}}

\newcommand{\LCperp}{{\scriptscriptstyle \perp}}

\newcommand{\Eins}{\mathbbmss{1}}

\newcommand{\A}{C}

\newcommand{\bw}{\begin{widetext}}
\newcommand{\ew}{\end{widetext}}

\newcommand{\fsl}[1]{\slashed{#1}}

\newcommand{\K}{K}

\begin{document}

\title{Scattering in plane-wave backgrounds: infra-red effects and pole structure}

\author{Anton Ilderton}
\email[]{anton.ilderton@physics.umu.se}
\affiliation{Department of Physics, Ume\aa\ University, SE-901 87 Ume\aa, Sweden}

\author{Greger Torgrimsson}
\email[]{greger.torgrimsson@physics.umu.se}
\affiliation{Department of Physics, Ume\aa\ University, SE-901 87 Ume\aa, Sweden}

\begin{abstract}
We consider two aspects of scattering in strong plane wave backgrounds. First, we show that the infra-red divergences in elastic scattering depend on the structure of the background, but can be removed using the usual Bloch-Nordsieck approach. Second, we analyse the infinite series of shifted-mass-shell poles in the particle (Volkov) propagator using lightfront quantisation. The complete series of poles is shown to describe a single, on-shell, propagating particle.
\end{abstract}
\pacs{}
%\tableofcontents
%
\maketitle
\section{Introduction}
Strong external fields affect many aspects of gauge theories. In QCD, magnetic fields affect the vacuum \cite{Chernodub:2011mc}, phase diagram \cite{Bali:2011qj}, electric dipole moments \cite{Basar:2011by} and the quark-gluon plasma \cite{Tuchin:2012mf}. In QED, scattering processes in the fields of intense lasers currently attract quite some interest \cite{Heinzl:2008an, DiPiazza:2011tq}, with the aim of investigating both nonperturbative effects~\cite{Dunne:2008kc} and beyond-standard-model physics~\cite{Jaeckel:2010ni,Redondo:2010dp}. Modelling the laser as a plane wave allows scattering amplitudes to be calculated for arbitrarily strong fields because the fermion propagator in a plane wave is known exactly; this is the Volkov propagator~\cite{Volkov}.

In this paper we will consider the propagation of quantum particles in strong plane waves. While the basic results for scattering in plane waves were given in the 1960s \cite{Reiss1, Nikishov:1963,Narozhnyi:1964}, those calculations assumed either monochromatic waves or crossed fields (constant plane waves). Both of these fields are of infinite extent and are therefore rather special cases; two statements related to them will be examined below.

We begin by considering the infra-red (``IR") structure of processes in strong plane waves, focussing on soft corrections to elastic scattering. In a crossed field, the differential probability of photon emission scales like $1/\omega^{2/3}$ rather than as  $1/\omega$ as in bremsstrahlung. Thus the logarithmic divergence of the IR becomes an integrable singularity. That this happens in a field which never vanishes is potentially interesting because IR problems originate in the (incorrect) assumption that the QED coupling switches off at large distances \cite{Kulish:1970ut,Horan:1999ba}. This weakening of the divergence actually comes at the expense of admitting unphysical large distance behaviour, see \cite{Dinu:2012tj}, but it has raised the question of whether a partially nonperturbative treatment of plane wave backgrounds can offer insight into the IR problem \cite{Bloch,Kinoshita:1962ur,Lee}, which is is still under active investigation \cite{Lavelle:2005bt,Lavelle:2009zz,Lavelle:2010hq, deOliveira:2012qa, Kitamoto:2011yx,Miao:2012xc}. It is also possible that background fields lead to new problems; it has been suggested, for example, that IR divergences do not factorise in pair-creating backgrounds \cite{Akhmedov:2009vh}. 

In the second part of the paper, we turn to the basic building block of elastic scattering, the electron propagator. Our focus is on the poles of the propagator, particularly in monochromatic fields. These are still used as an intuitive basis for more general calculations in finite pulses, and this has led to some debate concerning the `intensity dependent mass shift' \cite{Fried:1964zza, 49713} and \cite{debate1,debate2}. The mass shift (or rather, its effect) can be seen in the spectrum of undulator radiation \cite{Harvey:2012ie}, but its theoretical definition is more elusive. When it appears, the mass shift leads to poles in the propagator away from $p^2=m^2$, suggesting the presence of heavy states. We investigate this by directly constructing the quantum states of a particle in a plane wave and also by resumming the pole contributions to scattering amplitudes.

The paper is organised as follows. After a brief review of previous results, we discuss in Sect.~\ref{RESULTAT} loop and soft emission corrections to elastic scattering in plane wave backgrounds, and the cancellation of IR divergences following \cite{Yennie:1961ad,Weinberg}. In Sect.~\ref{LF} we discuss the propagator and the quantum states using lightfront quantisation, and explicitly relate the poles of the propagator to the ordinary mass shell condition.  We conclude in Sect.~\ref{CONCS}. Appendix \ref{APP0} collects useful results on the propagator and the normalisation of $S$-matrix elements in external plane waves. Appendix \ref{APP-IR} contains the details of our IR calculations.

\subsection{Conventions and review}
Consider a classical particle in a plane wave $F_{\mu\nu}(\phi)$ depending on $\phi=k.x$ with $k^2=0$, lightlike. We take $k.x=\omega x^\LCp$, lightfront time. (Recall that $x^\LCpm=x^0\pm x^3$, $x^\LCperp=\{x^1,x^2\}$, and $x^{\LCpm}=2 x_{\LCmp}$.) We consider finite duration fields, for which $F_{\mu\nu}$ vanishes before some $\phi_i$ and after some $\phi_f$. The lightfront structure of the plane wave means that {\it all} initially present particles enter (leave) the wave at the same lightfront time $\phi_i$ ($\phi_f$). A particle entering the wave with momentum $p_\mu$ has a subsequent kinematic momentum $\pi_\mu$ given by
\be\label{CL-SOL}
	\pi_\mu(p;\phi) = p_\mu - e \A_\mu(\phi) + k_\mu\frac{2ep.\A(\phi)-e^2 \A^2(\phi)}{2k.p}  \;,
\ee
in which $\A_\mu(\phi)$ is the integral of the electric field strength from the initial to the elapsed lightfront time:
\be\label{C-DEF}
	 \A_j(\phi) = \frac{1}{\omega}\int\limits_{\phi_i}^{\phi}\!\ud\varphi\ E^j(\varphi) \;, \quad j\in\{1,2\} \;,
\ee
and $\A_{\LCpm}=0$. No gauge potential is employed in this derivation, see \cite{Heinzl:2008rh}. We now set $\phi_i=0$. We are interested in `unipolar' pulses for which the integral over the entire electric field is nonzero \cite{unipolar2}, and which can be taken as crude models for fields which provide vacuum acceleration \cite{observed,Esarey:1995cw,Salamin:2002gh}. This means that after $F_{\mu\nu}$ has switched off, $\A^\mu$ becomes constant, and we can identify it with its value at $\phi=\infty$, so we write $\A^\mu(\phi) = C^\mu_\infty$ for $\phi>\phi_f$. Using (\ref{CL-SOL}), we then see that the difference in momentum for an electron passing through the wave obeys $(\pi(p;\infty)-p)^2 = e^2\A_\infty^2 \leq 0 \;,$ which has the right sign for scattering. When $C_\infty=0$, there is no net acceleration and $\pi(p;\infty)=p$. Note that $C_\infty$ is equal to the Fourier zero mode of the field strength~\cite{Dinu:2012tj}. 

Regarding gauge invariance in theories with plane wave backgrounds, there is nothing to discuss classically: particle motion is described by the Lorentz, LAD or LL equations which depend only on $F_{\mu\nu}$. Quantum mechanically, one must choose a potential. For plane waves, the almost universal choice is to take (up to a constant)
\be
	eA_\mu^\text{background} = e\A_\mu(\phi) \;, 
\ee
with $\A_\mu$ from (\ref{C-DEF}). This choice, often made implicitly, makes the physics manifest. Now that the gauge is fixed, final probabilities will depend on $\A_\mu$. We will see this in the appearance of $\pi_\infty :=\pi(p;\infty)$ in the probabilities below. Using (\ref{C-DEF}), though, we can always rewrite such results in terms of $F_{\mu\nu}$, which secures gauge invariance. (This explains the apparent result that quantum processes in plane waves ``depend on the vector potential characterising the laser field, not on the electric field component" \cite{Krajewska:2012qh}.)

\section{The infra-red sector}\label{RESULTAT}
\begin{figure}[t!]
	\includegraphics[width=0.34\columnwidth]{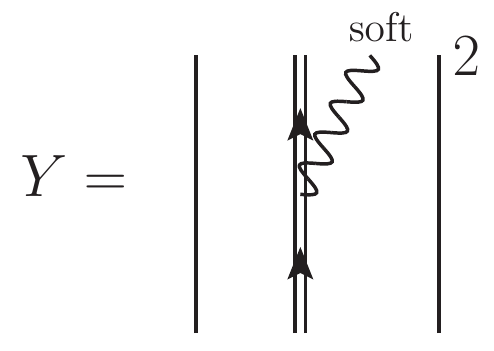}
	\caption{\label{Y-FIG} Nonlinear Compton scattering of a soft photon, at tree level. Double lines indicate the background-dressed Volkov propagator.}
\end{figure}

We now consider the infra-red behaviour of scattering processes in plane-wave backgrounds, through the use of several examples. The details and derivation of all our IR results may be found in Appendix B. We work in the Furry picture, in which the background field is treated exactly (i.e.\ without recourse to perturbation theory). We begin by recalling that the {\it soft contribution} to the probability of single photon emission, or `nonlinear Compton scattering'  see Fig.~\ref{Y-FIG} and \cite{Hartemann-review}, is \cite{Dinu:2012tj} 
\be\label{Y-DEF}
	Y :=-e^2\int\frac{\ud^dl}{(2\pi)^d}\frac{1}{2l_0}\left(\frac{\pi_\infty}{l.\pi_\infty}-\frac{p}{l.p}\right)^2 \;,
\ee
using dim reg in $d>3$ dimensions, and $Y$ is understood to carry an upper cutoff corresponding to, say, detector resolution. This is logarithmically divergent in $d=3$ when $\A_\infty\not=0$, but vanishes when $\A_\infty=0$. Hence, the presence of this IR divergence depends on the properties of the background. 

IR divergences typically arise when virtual particles come close to the mass shell. The field-dependent IR divergences in plane-wave backgrounds arise as follows. At each vertex, the structure of the background allows the $x^\LCm$ and $x^\LCperp$ integrals to be performed immediately. The $p_\LCp$ integral can be performed using the residue theorem, which restricts the remaining $x^\LCp$ integral by introducing a lightfront time-ordering. One then sees that it is the large lightfront time parts of these integrals which yield singularities. In other words, our IR divergences essentially arise from the background-free regions of spacetime, before and after the pulse. See Appendix~B.

%%%%%%%%%%%%%%%%%%%%%%%%%%%
%%%%%%%%%%%%%%%%%%%%%%%%%%%
\subsection{Double Compton scattering: hard-soft factorisation}
%%%%%%%%%%%%%%%%%%%%%%%%%%%
%%%%%%%%%%%%%%%%%%%%%%%%%%%
%
Consider now the emission of two photons from an electron in a plane wave, i.e.\ the process
\be
	e^-(p) \overset{\text{in laser}}{\to} e^-(p') + \gamma(k') + \gamma(l) \;,
\ee
 as recently investigated in \cite{Lotstedt:2009zz,Seipt:2012tn, Mackenroth:2012he}. Assume that the photon with momentum $l_\mu$ is soft. It can be emitted from either the incoming or outgoing leg, and the $S$-matrix element takes the form
\be\label{FACTOR}
	S_{fi} = e \,\epsilon_\text{soft}.\bigg(\frac{p}{l.p}-\frac{p'}{l.p'}\bigg) S_{NLC}(p\to p',k') \;,
\ee
in which $S_{NLC}$ is the $S$-matrix element for nonlinear Compton.  This is the expected form of a soft correction to a hard scattering process;  two-photon emission becomes degenerate with nonlinear Compton when one of the emitted photons is soft. The soft divergence implied by (\ref{FACTOR}) is {\it independent} of the structure of the background. This is an example of a general result (see the appendix): the structure of the plane wave has no impact on the hard-soft factorisation of IR divergences, or the severity of those divergences, which give the usual $1/\varepsilon_\text{\tiny IR}$ poles in $4+2\varepsilon_\text{\tiny IR}$ dimensions. (See \cite{Morozov-photo,Morozov-elastic} for examples in crossed fields.) Essentially, taking the soft limit removes the background-field dressing from emission vertices, and factorisation proceeds as in QED without background.

However, the soft divergence implied by (\ref{FACTOR}) is not the highest order divergence in two-photon emission. Rather, this comes from the case in which both photons are soft; two-photon emission then becomes becomes degenerate with elastic scattering, see Fig.~\ref{1}. The IR divergent part of the emission probability in this case is
\be\label{SISTA-P}
	\mathbb{P}\overset{\text{\tiny IR}}{=}\frac{1}{2}\bigg[e^2\! \int\! \frac{\ud^d l}{(2\pi)^d 2 l_0} \bigg(\frac{\pi_\infty}{l.\pi_\infty}-\frac{p}{l.p}\bigg)^2\bigg]^2 = \frac{1}{2}Y^2\;.
\ee 
Unlike in (\ref{FACTOR}), this divergence does depend on the structure of the background. For $\A_\infty\not=0$ each integral contributes the same divergent term; the leading singularity is therefore $1/\varepsilon_\text{\tiny IR}^2$.

\subsection{Elastic scattering}
\begin{figure}[t!]
	\includegraphics[width=0.35\columnwidth]{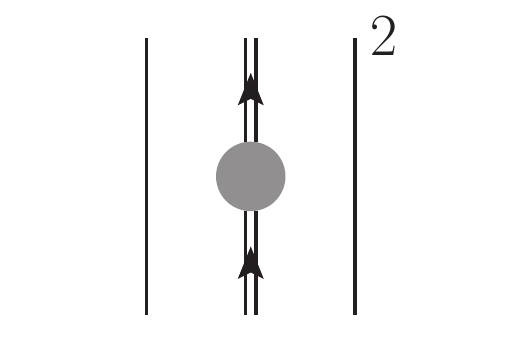}
	\caption{\label{1} The probability for elastic scattering. The grey dot denotes all loop corrections.}
\end{figure}
As shown in Appendix B (leading to (\ref{X-derive})), the elastic scattering probability including the soft contribution from all loop orders, see Fig.~\ref{1}, is
\be\label{ALLA-LOOPAR}
\mathbb{P}\overset{\text{\tiny IR}}{=}\exp\bigg[e^2\int\frac{\ud^dl}{(2\pi)^d}\frac{1}{2l_0}\left(\frac{\pi_\infty}{l.\pi_\infty}-\frac{p}{l.p}\right)^2\bigg] =: e^X \;,
\ee
which defines $X$. This expression has two parts. The exponential is the all-orders soft loop contribution which is log divergent in $d=3$ when $\A_\infty\not=0$. This multiples a `1' which is, see the appendix, the exact tree-level probability of elastic scattering. (That $\mathbb{P}=1$ at tree level is already a sign that something is wrong.) 

These problems are related and their resolution is clear: it is not possible to observe `elastic scattering' alone, due to the potential emission of soft, unobservable, photons. When we calculate the inclusive probability, IR divergences coming from soft emission should cancel those coming from the loops.  We turn to this now.

\subsection{IR cancellation}
\begin{figure}[t!]
	\includegraphics[width=\columnwidth]{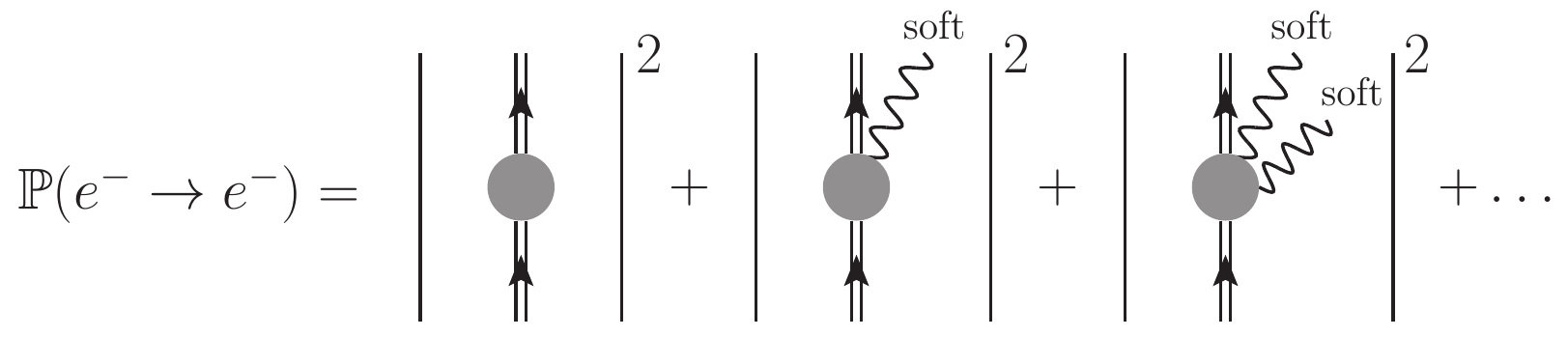}
	\caption{\label{SISTA} The measurable probability of `scattering without emission' is the sum of the probabilities for elastic scattering, and the probabilities for the emission of arbitrary numbers of soft (unobserved) photons.}
\end{figure}

%%%%%%%%%%%%%%%
%
Generalising (\ref{SISTA-P}), the tree level probability of emitting $n$ soft photons is $Y^n/n!$. The all orders soft-loop contribution to each of these processes is $e^X$, as in (\ref{ALLA-LOOPAR}). What can be observed in an experiment is the sum of 1) the probability for elastic scattering and 2) the probabilities for emission of any number of undetected soft photons. This sum, see Fig.~\ref{SISTA}, is the measurable probability of observing scattering of the electron without photon emission, and its IR part is 
\be\begin{split}
	\mathbb{P}(e^\LCm \to e^\LCm) &\overset{\text{\tiny IR}}{=} e^X \cdot 1 + e^X \cdot (Y + \tfrac{1}{2}Y^2 +\ldots) \\
	&\overset{\text{\tiny IR}}{=} e^{X+Y}\;.
\end{split}
\ee
We see that the IR  contributions factorise. Comparing (\ref{Y-DEF}) and (\ref{ALLA-LOOPAR}), we see also that $X=-Y$, so that
\be\begin{split}
	\mathbb{P}(e^\LCm \to e^\LCm) &\overset{\text{\tiny IR}}{=} 1 \;,
\end{split}
\ee
and the leading field-dependent soft divergences cancel to all orders. Two remarks are now in order.

First, we have followed \cite{Weinberg} and considered only the divergent IR contributions to amplitudes, showing that these cancel. In order to extend our results to the complete amplitudes, i.e.\ in order to include the IR finite parts, one can instead follow the method of~\cite{Yennie:1961ad}. (See \cite{MITTER, Morozov-photo, Morozov-elastic,Meuren:2011hv} for various loop calculations.)

Second, there are no purely soft divergences in a crossed field \cite{Nikishov:1963}. In such a field, all particles are accelerated to the speed of light. Further, the literature results assume that the particle is also {\it initially} moving at the speed of light. This leads to the replacements $\{\pi,p\}\to k$ in (\ref{SISTA-P}) and (\ref{ALLA-LOOPAR}), so that $X$ and $Y$ vanish individually. If a field which is nonzero and constant for a long but finite time is used instead, the $\A_\infty$ dependent logarithmic divergence reemerges \cite{Dinu:2012tj}. Thus, the `improved' IR behaviour in crossed fields comes at the expense of introducing unphysical large distance behaviour. It does not seem to say anything about the large distance structure of QED.

%
%%%%%%%%%%%%%%%%%%%%%%%%%%%
%%%%%%%%%%%%%%%%%%%%%%%%%%%
\subsection{Indistinguishable processes}
%%%%%%%%%%%%%%%%%%%%%%%%%%%
%%%%%%%%%%%%%%%%%%%%%%%%%%%
%
Suppose now that $\A_\infty=0$, i.e.\ there is no vacuum acceleration, and consider again nonlinear Compton scattering. The final state integrals are now IR finite, and one can integrate over all photon momenta to calculate the `full probability'. This number, though, is not measurable. Even when processes are IR finite, one must still account for indistinguishable processes. 

Physically, this issue originates in the nonzero frequency range of a pulsed field. If we parameterise this frequency range as $s\omega$ where $s$ is real, then each $s$ can produce photons with a frequency $\omega'_s$, where (taking the incoming electron to be at rest for simplicity) \cite{Dinu:2012tj}
\be\label{OMEGA-NLC}
	\frac{1}{\omega'_s} = \frac{1}{s\omega} + \frac{1}{m}(1-\cos\theta) \;.
\ee
Each $\omega'_s$ is bounded below, $\omega'_s> \omega s$, but this lower bound can extend down to zero frequency in a pulse, even if there is no support at zero frequency itself, i.e.\ if $\A_\infty=0$ \cite{Dinu:2012tj}.  Suppose, then, that an experiment can detect only photons of frequency higher than $\omega_0$. The measurable probability for one-photon emission is then the sum of that for nonlinear Compton scattering with $\omega'>\omega_0$, together with the sum of all $n$-photon emission probabilities in which $n-1$ photons are soft, $\omega'<\omega_0$.

For a discussion of indistinguishable processes and the structure of the Volkov propagator, see \cite{MLDM}. The essential part of the amplitudes calculated above is in fact just this propagator: elastic elastic at tree level is given by the double amputation of the propagator, see Appendix \ref{APP0}, and this is multiplied at higher orders by soft corrections. We therefore turn now to the propagator itself.

%%%%%%%%%%%%%%%%%%%%%%%%%%%
%%%%%%%%%%%%%%%%%%%%%%%%%%%
\section{The poles of the propagator} \label{LF}
%%%%%%%%%%%%%%%%%%%%%%%%%%%
%%%%%%%%%%%%%%%%%%%%%%%%%%%
%%
%
The K\"all\'en-Lehmann representation of the two-point function provides a natural way to identify the mass of single-particle states, as the location of the poles in momentum space \cite{KL}. One of the most discussed properties of the Volkov propagator is its infinite series of poles at
\be\label{quasi-poles}
	(p-l k)^2 = m_*^2 \;,\quad l\in\mathbb{Z}\;,
\ee
where $m_*^2 = m^2(1+a_0^2)$ is the shifted mass of a particle in a plane wave of intensity $a_0$ \cite{Sengupta,Sarachik:1970ap, Heinzl:2008rh}. There is no pole at the ordinary mass shell. The $l=0$ pole, $p^2 = m_*^2$ might suggest a change in the rest mass, while the other poles describe such a heavy particle absorbing `multiple photons' from the background. The literature contains many interpretations of the poles. They were originally considered to be problematic \cite{Oleinik}, but that higher loop corrections would regulate the poles by giving them a finite height and width \cite{Becker:1976ne}. The fact that the poles are discrete has lead to the claim that the spectrum of a particle in a plane wave is discrete \cite{Oleinik,EKK}. See \cite{Eberly} for an interpretation of the mass-shift $m_*$ as a finite mass renormalisation, and \cite{Bergou:1980du} for an interpretation in terms of effective potentials.

Our focus is on this pole structure. Spin effects do not impact our discussion so we restrict to a scalar particle. Propagator poles are usually tied to the optical theorem and intermediate states, so we will begin by constructing the quantum states of a particle in a plane wave explicitly. We will see that loop corrections are not, in fact, necessary, hence we turn off the QED interaction. We therefore consider only a scalar particle in a plane wave; since this background cannot spontaneously produce pairs \cite{Schwinger:1951nm}, particle number is conserved and the theory is simple. However, the equal-time quantisation of this theory does not seem to appear in the literature. The principle difficulty lies in proving certain orthogonality relations of the Volkov solutions \cite{RitusReview,BOCA}.  The reason for this difficulty is that the background singles out preferred {\it lightlike} directions. It is therefore natural to tackle problems in plane waves using lightfront quantisation, as suggested in \cite{Neville:1971uc}. We will show here that the theory is trivially quantised on the lightfront. We will then recover the Volkov propagator as a lightfront time-ordered product and investigate the poles. The explicit equal-time quantisation of the theory will be performed in the forthcoming article~\cite{MLDM}.

\subsection{Particle states in plane waves.}
A complex scalar in a plane wave obeys the equations of motion $(D^2+m^2)\varphi=0$. The general solution is
\be\label{GenSol}
	\varphi(x) = \int\!\frac{\ud^3{\sf p}\,\theta(p_\LCm)}{(2\pi)^3 2p_\LCm}	
\  a_{\sf p}\varphi_{\sf p}(x) + b^\dagger_{\sf p}\varphi_{-\sf p}(x) \;, \\
\ee
where ${\sf p}:= \{ p_\LCperp, p_\LCm\}$. The functions $\varphi_{\sf p}$ are scalar Volkov solutions obeying the initial condition $\varphi_{\sf p}(\phi=0)= e^{-i{\sf p}.\sf{x}}$. The set $\{a_{\sf p}, b_{\sf p}\}$ is therefore the initial data at $\phi=0$, and the Volkov solutions recover the classical kinematic momenta of a particle in a plane wave via
\be\label{GET-KINETIC}
	iD_\mu(\phi) \varphi_{\sf p}(x)= \pi_\mu(p;\phi)\varphi_{\sf p}(x) \;.
\ee
To quantise, we impose the lightfront commutation relation (LCR) \cite{Heinzl:2000ht},
\be\label{CCR0}
	\big[\varphi(x),2\partial_\LCm\varphi^\dagger(y)\big]\big|_{x^\LCp = y^\LCp} = i\delta^{\LCperp,\LCm}(x-y) \;.
\ee
Using (\ref{GenSol}) to calculate the left hand side of (\ref{CCR0}), the complicated exponentials in the Volkov solutions cancel immediately, and the commutator reduces to that of the free lightfront theory. The LCR is obeyed if
\be\label{MODE-CCR}
	[a_{\sf p},a^\dagger_{\sf q}] =  [b_{\sf p},b^\dagger_{\sf q}]  = (2\pi)^32p_\LCm\delta^3({\sf p}-{\sf q}) \;,
\ee
which are the usual free-field commutators on the lightfront. The particle interpretation of our theory is as follows. We define the lightfront vacuum $\ket{0}$ to be annihilated by the $a_{\sf p}$ and $b_{\sf p}$ as usual, and then the first excited states are
\be\begin{split}\label{THE-STATES}
	\ket{p} := a^\dagger_{\sf p}\ket{0} \;,\quad\text{and}\quad \ket{\bar p} := b^\dagger_{\sf p}\ket{0} \;,
\end{split}
\ee
which we will now see are one-particle/antiparticle states respectively, just as in the free theory. Using the energy momentum tensor, we combine the normal ordered Hamiltonian and kinematic momenta into 
\be\begin{split}
	&\Pi_\mu(\phi) = T_{-\mu} = \int\!\ud^3{\sf x}\ \partial_\LCm\varphi^\dagger D_\mu\varphi(x) + \text{c.c.\ ,} \\
	&= \int\!\frac{\ud^3{\sf p}\,\theta(p_\LCm)}{(2\pi)^32p_\LCm}\  \pi_\mu(p; \phi) a^\dagger_{\sf p}a_{\sf p} + \bar{\pi}_\mu({p}; \phi) b^\dagger_{\sf p}b_{\sf p}  \;,
\end{split}
\ee
where $\bar\pi_\mu\equiv \pi_\mu\big|_{e\to -e}$. The states (\ref{THE-STATES}) then have time-dependent energies and momenta given by 
\be
\begin{split}\label{EIGEN}
	\Pi_\mu(\phi)\ket{p} &= \pi_\mu( p;\phi)\  \ket{ p} \;, \\
	\Pi_\mu(\phi) \ket{\bar p} &= \bar\pi_\mu( p;\phi)\  \ket{\bar p} \;.
\end{split}
\ee
The excited states therefore carry the time-dependent momenta of the classical theory. So, $a^\dagger$ and $b^\dagger$ create particles which, on the initial surface $\phi=0$, have on-shell kinetic momenta $p_\mu$. This labels the {\it continuous} spectrum of states. At all subsequent times, the states carry momentum $\pi_\mu({p}; \phi)$ with $\pi^2=m^2$.

%%%%%%%%%%%%%%%%%%%%%%%%%%%%
%%%%%%%%%%%%%%%%%%%%%%%%%%%%
\subsection{The propagator}\label{PROP-SECT}
%%%%%%%%%%%%%%%%%%%%%%%%%%%%
%%%%%%%%%%%%%%%%%%%%%%%%%%%%
%
We have seen that the one-particle states have mass $m^2$, and that this spectrum of states is continuous. The background-dependent structure in the field operators (essentially the Volkov solutions) simply describes the action of the Lorentz force on a particle in a background field. How should this be reconciled with (\ref{quasi-poles}) which might suggest that additional ``heavy" states appear? To answer this we turn  to the propagator, which is the lightfront time-ordered product of the fields (\ref{GenSol}):
\be\begin{split}\label{T-PROD}
	G(x,y) &= \bra{0} \mathcal{T}_\LCp \varphi(x)\varphi^\dagger(y)\ket{0} \\
&= \int\!\frac{\ud^3{\sf p}\,\theta(p_\LCm)}{(2\pi)^32p_\LCm} \theta(x^\LCp-y^\LCp)\varphi_{\sf p}(x)\varphi^\dagger_{\sf p}(y) \\
		&\hspace{60pt}+ \theta(y^\LCp-x^\LCp)\varphi_{\LCm\sf p}(x)\varphi^\dagger_{\LCm\sf p}(y) \;.
\end{split}
\ee
The lightfront time ordering may be made covariant by using an $i\epsilon$ prescription to yield
\be\label{VPROP}
\begin{split}
	G(x,y) &=i \int\!\frac{\ud^4 p}{(2\pi)^4} \frac{e^{-ip.(x-y)-\frac{i}{2k.p}\int\limits_{k.y}^{k.x} 2eA.p-e^2A^2}}{p^2-m^2+i\epsilon} \\
	&=  i \int\!\frac{\ud^4 p}{(2\pi)^4} \frac{\varphi_{p}(x)\varphi^\dagger_{p}(y)}{p^2-m^2+i\epsilon}  \;.
\end{split}
\ee
This is the Volkov propagator \cite{Brown:1964zzb, Kibble:1975vz}.  In the second line, we have written the Volkov solutions $\varphi_p(x)$ with a script label to indicate that $p_\mu$ is arbitrary, i.e.\ $p^2\not=m^2$ in general. Performing the integral over $p_\LCp$ (on which $\varphi_p$ depends trivially), the simple pole returns us to (\ref{T-PROD}), putting the initial momentum $p_\mu$, and therefore the kinetic momentum $\pi_\mu$, onto the mass-shell.

Since we can write $p^2-m^2 \equiv \pi^2(p;\phi)-m^2$, we see that (\ref{VPROP}) is a spectral representation of the explicitly time-dependent operator $D^2+m^2$. It is not, in contrast to the free propagator, a Fourier representation, which is available only for simple fields such as monochromatic waves. Fortunately, this is just the case of interest. So, consider a a circularly polarised, monochromatic field
\be
	\A_\mu(\phi) = l^1_\mu\sin \phi+l^2_\mu \cos \phi\;, 
\ee
with $e^2 l^i.l^j = -m^2a_0^2\delta^{ij}$, $l^j.k=0$. The intensity $a_0$ appears here and is equal to $eE_{rms}/m\omega$. We define 
\be
	r e^{i\theta} =i\frac{l^1.p}{k.p}+\frac{l^2.p}{k.p}\quad \text{and}\quad q_\mu = p_\mu+\frac{a_0^2}{2k.p}k_\mu \;.
\ee
The well-known quasimomentum $q_\mu$ obeys $q^2=m^2(1+a_0^2)\equiv m_*^2$, yielding the shifted mass. The Volkov solutions in this field are
\be\label{circular Volkov}
	\varphi_p(x)=e^{ -iq.x-ir\sin(\phi-\theta)-ir\sin(\theta)} \;.
\ee
(The final term in the exponent follows from initial conditions, but is usually dropped in the literature. Our results hold in either case.) The Fourier transformed propagator can be constructed directly from (\ref{T-PROD}) or (\ref{VPROP}) and coincides with the known result \cite{Brown:1964zzb, Kibble:1975vz,Eberly}
\be\label{PROP-MON}
\begin{split}
	{\tilde G}_\text{mon}&(p',p) = \sum\limits_{n,l\in\mathbb{Z}}   J_{n+l}\big(r)J_l(r) e^{i n\theta}\\
	&(2\pi)^4\delta^4(p'-p-nk)\frac{i}{(p-lk)^2-m_*^2+i\epsilon}  \;.
\end{split}
\ee	
The delta-comb structure is due to the periodicity of the background \cite{Schwinger:1949ym,Eberly}. Since the Bessel functions are everywhere regular, we see that when $p_\mu$ and $p'_\mu$ are such that the $n^\text{th}$ delta-function has support, we recover the infinite series of poles (\ref{quasi-poles}).

\subsection{Pole resummation}

Writing a Feynman amplitude in momentum space, internal lines become the Fourier transform of the propagator. From (\ref{PROP-MON}), we see that one then hits a ``resonance" for particular values of $l$ and $n$, each of which corresponds to taking particular values of energy-momentum from the background\footnote{Related structures are seen in the lower order three-point processes of strong-field QED; due to the periodicity of the background, the emission rates take the form of a diffraction pattern. When the momentum transfer over a cycle is a multiple of the laser frequency, there is a peak in the emission rate which is analogous to a patch of constructive interference \cite{Neville:1971uc, Heinzl:2010vg}.}. In general, all the poles terms contribute to a given amplitude, and there is no obvious way to single out a particular pole over any other. We therefore consider the sum of contributions from all the poles. This can be extracted by taking  real and imaginary parts, i.e.\ applying the standard result
\be\label{REandIm}
	\frac{1}{x+i\epsilon} = -i\pi \delta(x) + \frac{\mathcal{P}}{x} \;,
\ee
to (\ref{PROP-MON}) and retaining only the delta function terms; call this part of the propagator $R$. In position space, one has
\begin{widetext}
\be
\begin{split}
	R(x,y) &= \int\frac{\ud^4(p,p')}{(2\pi)^8}e^{ip.y+ip'.x}\sum\limits_{n,l\in\mathbb{Z}} J_{n+l}(r)J_l(r)e^{in\varphi}\, (2\pi)^4\delta(p'-p-nk)\, \pi\delta((p-lk)^2-m_*^2) \;.
\end{split}
\ee
\end{widetext}
Performing the $p'$ integrals, changing variables $n\rightarrow s=n+l$ and $p\rightarrow p-lk+\frac{a^2}{2kp}k$, and summing over the Bessel functions using
\be
\sum\limits_{s\in\mathbb{Z}} J_s(r)e^{-is(k.x-\varphi)}=\exp\big(-ir\sin(kx-\varphi) \big) \;.
\ee
gives the final result
\be\label{RDEF}
	R(x,y)=\int\frac{\ud^4p}{(2\pi)^4}\pi\delta(p^2-m^2)\varphi_p(x)\varphi_p^*(y) \;,
\ee
which is a sum over on-shell Volkov wavefunctions (\ref{circular Volkov}). The same result is obtained directly from (\ref{VPROP}) by sending 
\be
	\frac{i}{p^2-m^2+i\epsilon}\rightarrow\pi\delta(p^2-m^2) \;.
\ee
Hence, the total contribution of the infinite series of poles is to replace the propagator by an integral over all real, on-shell intermediate states. In other words, the poles contribute the `imaginary parts' one obtains from cutting the propagator, in the sense of the optical theorem. This confirms that the off-shell poles do not describe heavy states.  The reason that the usual K\"all\'en-Lehmann interpretation does not go through directly is that its derivation assumes Poincare invariance of the theory. This is explicitly broken by the presence of background fields (though covariance is not). Related to this, the Fourier transform cannot be interpreted in the same way as in a free theory since canonical momentum (the Fourier variable) and kinematic momentum are not equal. See \cite{Dusedau:1985ue} for analogous statements and an investigation of the K\"all\'en-Lehmann representation in AdS space.

Our treatment of the poles has been formal: the poles do play a role in the detailed structure of emission rates \cite{Seipt:2012tn}, and they collectively describe the regime of momentum exchange in which sufficient energy is taken from the background to put normally virtual (intermediate) particles onto the mass-shell. Associating individual poles to physical states is misleading, though. The confusion arises because in monochromatic (periodic) waves, the one-particle states are also eigenstates of cycle-averaged momentum operators, with constant eigenvalues equal to the quasi-momenta, which square to the shifted mass. One then speaks of the quasi-momenta as the `good quantum numbers' of the system \cite{Zel}. However, such nonlocal operators do not tell us much about the states. A particle in a background field represents a time-dependent problem and so `eigenvalues' are in general time-dependent, as in (\ref{EIGEN}).

The results above extend to general plane waves as follows. Given Volkov solutions $\varphi_p(x)$ in a particular plane wave, we define $R(x,y)$ as in (\ref{RDEF}) and introduce the Fourier transform $\Gamma$ via
\be
		\varphi_p(x) = e^{-ip.x}\int\!\frac{\ud s}{2\pi}\ e^{-isk.x}\ \Gamma_s(p) \;,
\ee
from which we obtain the implicit Fourier transform
\begin{widetext}
\be\begin{split}\label{THE-FT}
	\tilde{G}(p',p) 
 	&=\int\!\frac{\ud l\, \ud n}{(2\pi)^2} \ \Gamma_{n+l}(p)\, \Gamma^*_{l}(p)\  (2\pi)^4\delta^4(p'-p-nk)\frac{i}{(p-lk)^2-m^2+i\epsilon} \;.
\end{split}
\ee
It is then trivial to check that
\be
	\tilde{R}(p',p) = \int\!\frac{\ud l\, \ud n}{(2\pi)^2} \ \Gamma_{n+l}(p)\, \Gamma^*_{l}(p)\  (2\pi)^4\delta^4(p'-p-nk)\ \text{Re}\bigg[\frac{i}{(p-lk)^2-m^2+i\epsilon} \bigg]\;.
\ee
The sum over on-shell intermediate states, $R$, is therefore given by taking the real part of the free propagator buried inside $G$. In this context the contributions from the background act as a `dressing' which encodes the time dependence of the system, but the plane wave does not change the fundamental particle content of the theory.

\end{widetext}

\section{Conclusions}\label{CONCS}
Hard-soft factorisation of scattering processes in QED goes ahead in the presence of a plane wave background field of arbitrary strength and shape. The factorisation is not sensitive to the structure of the background. The implication is that, just as in ordinary QED, IR divergences in plane wave backgrounds exponentiate and cancel from measurable processes. A rough explanation of why is as follows. A scattering process in a fixed background obeying Maxwell's equations can always be rewritten as scattering between asymptotic, coherent, photon states. As these are free-theory states, the scattering process is equivalent to a sum over ordinary QED process with all numbers of photons. Hence, if the IR divergences cancel in QED, they should also cancel here. (Compare \cite{Akhmedov:2009vh}, which suggests that pair-creating backgrounds not obeying Maxwell may lead to non-factorisable divergences.)

However, in those processes which are entirely soft, e.g.\ elastic scattering, the IR divergences depend on the Fourier zero mode of the field strength (if this is nonzero, the pulse can transfer net energy to a classical particle passing through it). Nevertheless, we have shown that the soft IR divergences in loop corrections to elastic scattering are cancelled by divergences coming from multiple soft photon emissions, as normal.

The structure of the plane wave leads to the appearance of lightfront time-ordering in scattering amplitudes, and IR divergences then arise from the large lightfront time regions before and after the pulse. This suggests that the natural setting for strong field QED is lightfront quantisation, as employed in \cite{Neville:1971uc}. We used lightfront quantisation to explicitly constructed the quantum states of a particle in a plane wave (which are continuous, not discrete as previously claimed), and recover the Volkov propagator as a lightfront-time-ordered product. We showed that the shifted mass-shell poles in the Fourier representation of the propagator actually correspond to going onto the ordinary mass-shell. The reason that the poles do not correspond to particle masses is due to  the explicit breaking of Lorentz invariance induced by the background; the presence of the laser means that the Fourier variable does not coincide with physical momentum, and hence the K\"all\'en-Lehmann interpretation of the poles does not apply to the Fourier representation of the propagator.  It seems more natural, therefore, to talk of the mass-shift in terms of its observable effects, namely the spectral properties of photons emitted in nonlinear Compton scattering \cite{49713}. See  \cite{Harvey:2012ie} for a discussion of such effects beyond the monochromatic approximation.

\acknowledgements
The authors are grateful to A.~Signer for providing many helpful comments on a draft of this paper.  A.~I. thanks V.~Dinu, T.~Heinzl, M.~Lavelle and D.~McMullan for useful discussions, and D.~Seipt for providing references. The authors are supported by the Swedish Research Council, contract 2011-4221. Diagrams created using JaxoDraw \cite{Binosi:2003yf,Binosi:2008ig}.

\appendix

\section{The Volkov propagator and wavefunctions}\label{APP0}
\subsection{LSZ reduction}
We write $e\A(\phi)= a(\phi)$  from here on to compactify notation. We work in the Furry picture, treating the coupling to the background exactly and the interactions between the quantised fields in perturbation theory as normal. Feynman diagrams are therefore built from ordinary QED vertices and the spinor Volkov propagator $S$, which is the inverse of $\epsilon-i[i\slashed{D}-m]$:
\be\label{Propagator}
\begin{split}
	S(x,y)=&i\int\!\frac{\ud^4q}{(2\pi)^4}\ \K_{qx}\frac{e^{-iq.(x-y)}}{\fsl{q}-m+i\epsilon}\bar{\K}_{qy}\, e^{-i\int\limits_{k.y}^{k.x}V_q} \;,
\end{split}
\ee
where we have defined
\be
	\K_{px}:= \Eins+\frac{\fsl{k}\fsl{a}}{2k.p} \;, \quad V_p=\frac{2a.p-a^2}{2k.p} \;,
\ee
and $\bar\K = \gamma^0\K^\dagger \gamma^0$. For $S$-matrix elements we also need LSZ reduction, which, in the absence of background fields, tells us to replace external legs with free particle wavefunctions. In a background, LSZ reduction transforms external propagators into incoming and outgoing fermion wavefunctions; the following short calculation shows this explicitly in plane waves (although the result holds more generally). According to LSZ reduction, the incoming electron wave function is given by
\be\label{LSZ-0}
	\Psi_p^{\text{in}}(x)=-i\int\!\ud^4 y\ S(x,y) [-i\overset{\leftarrow}{{\fsl{\partial}}}_y-m]\, e^{-ip.y}u_p\;.
\ee
Since $\bar{\K}\overleftarrow{\fsl{\partial}}=0$, we find $-i\overleftarrow{\fsl{\partial}}\rightarrow\fsl{q}+V_q\fsl{k}$ in (\ref{Propagator}). The $y^\LCperp,y^\LCm$ integrals set ${q}_{\LCm,\LCperp}={p}_{\LCm,\LCperp}$. Writing $q=p+tk$ we get, after simplifying the spin term,
\bea
	\Psi_p^{\text{in}}(x) &=&\K_{px}u_p e^{-ip.x} \\
\nonumber	&& \int\!\ud\phi_y\int\!\frac{\ud t}{2\pi}\left(1+\frac{V_p}{t+i\epsilon}\right) e^{-it(\phi_x-\phi_y)-i\int\limits_{\phi_y}^{\phi_x}V_p} \;.
\eea
Now we perform the $t$-integral. The first term in the round brackets gives a delta function which sets $\phi_y=\phi_x$. The second term yields a step function, and the resulting $\phi_y$ integral is exact. Performing this integral we obtain
\be\label{VIN}
	\Psi_{p}^{\text{in}}(x)= \K_{px}u^\sigma_p \exp\bigg(-ip.x-i\int\limits_0^{\phi_x}V_p\bigg) \;,
\ee
which is the Volkov electron wavefunction (the solution to the Dirac equation in a plane wave) with kinetic momentum $p^{\mu}$ and spin $\sigma$ in the far past. Its current is
\be
	\frac{1}{2m}\bar\Psi_p^\text{in}\gamma_\mu \Psi_p^\text{in} = \pi_\mu(p;\phi) \;,
\ee
which corresponds to a particle with kinematic momenta (\ref{CL-SOL}). Similarly, see \cite{Dinu:2012tj}, the outgoing electron wavefunction which carries kinetic momentum $p^{\mu}$ and spin $\sigma$ in the far {\it future} is given by 
\be\label{VOUT}
	\bar{\Psi}_{p}^{\text{out}}(x)=\bar{u}_p^\sigma\, \delta\bar{\K}_{px}\exp\bigg(i(p+a_\infty).x-i\int\limits_{\phi}^{\phi_f}\delta V_p\bigg)   \;,
\ee
in which, and from here on, $\delta$ means $a\rightarrow a-a_\infty$. Positron solutions are obtained by sending $u\to v$ and $a\to -a$. In summary, LSZ transforms external lines into Volkov wavefunctions, which describe (on mass-shell) particles in a background plane wave.

\subsection{Normalisation}
The use of the Furry picture Feynman rules and the LSZ formulae above correspond to calculating $S$-matrix elements using equal-time quantisation as normal, with asymptotic momentum states $\ket{p}$ obeying 
\be\label{normalisation1}
\bracket{q}{p}=2p_0(2\pi)^3 {\delta}^3({\bf p}-{\bf q}) \;.
\ee
The free states evolve in time to become Volkov wavefunctions. The structure of the plane wave means firstly that these wavefunctions are naturally normalised on the lightfront (not as in (\ref{normalisation1})), and secondly that $S$-matrix elements of such wavefunctions conserve overall ${\sf p}:=\{p_\LCm, p_\LCperp\}$, not three-vector $\bf p$. We give here a clear prescription for dealing with normalisations which eliminates the need for volume factors or trying to compare the infinite volumes $\delta^3({\bf p})$ and $\delta^3_{\LCperp,\LCm}({\sf p})$, see also \cite{Boca:2009zz,Seipt:2010ya}.

An $S$-matrix element calculated using the Volkov solutions (\ref{VIN}) and (\ref{VOUT}) for an incoming electron, momentum $p_\mu$, and a set of outgoing particles with momenta $\{p_f\}$ takes the form (sums/products of $p_f$ are implicit) 
\be\label{SFI-GEN}
	S_{fi} =(2\pi)^3\delta^3_{\LCm,\LCperp}({\sf p}_f -{\sf p}) M(p\to p_f)\;,
\ee
which defines $M$. Now, we should really consider scattering between properly normalised wavepackets rather than momentum states. Final states will always be integrated out to obtain the full probability with Lorentz invariant measure
\be
	\sum\limits_{p_f} = \prod_f \int\!\frac{\ud^3 p_f}{(2\pi)^32p_{f0}} \;,
\ee
and it is enough to consider only the incoming electron wavepacket. This corresponds to multiplying (\ref{SFI-GEN}) by the factor
\be
	\int\!\frac{\ud^3{\bf p}}{\sqrt{(2\pi)^3 2p_0}} \psi(p) \quad\text{with}\quad \int\!{\ud^3{\bf p}}\ |\psi(p)|^2  = 1\;.
\ee
The $S$-matrix element mod-squared then becomes
\bea\label{S2-GEN2}
	|S_{fi}|^2 &=&\int\!{\ud^3{\bf p}}\ |\psi(p)|^2 \\
\nonumber	&&  (2\pi)^3\delta^3_{\LCm,\LCperp}({\sf p}_f -{\sf p}) |M(p\to p_f)|^2 \frac{1}{2p_-} \;.
\eea
Making the usual assumption that the wavepacket is sharply peaked corresponds to calculating (\ref{SFI-GEN}) and then dropping the first line in (\ref{S2-GEN2}) for $|S_{fi}|^2$. In short, the incoming electron should carry a normalisation factor of $1/2p_\LCm$, rather than the usual $1/2p_0$, at the level of the probability, the final expression for which is, summed (averaged) over final (initial) polarisation and spin
\be\label{absolutSISTA}
	\mathbb{P} = \frac{1}{2}\sum\limits_{p_f,\sigma,\epsilon} \frac{1}{2p_\LCm}(2\pi)^3\delta^3_{\LCperp,\LCm}({\sf p}_f-{\sf p})|M(p\to p_f)|^2 \;.
\ee
%
%%%%%%%%%%%%%%%%%%%%%%%%%
%%%%%%%%%%%%%%%%%%%%%%%%%
\subsection{Example: elastic scattering}\label{APP1}
%%%%%%%%%%%%%%%%%%%%%%%%%
%%%%%%%%%%%%%%%%%%%%%%%%%
%
As an example, consider the matrix element for elastic scattering at tree level. According to the LSZ reduction formulae this is obtained by amputating the Volkov wave function for the incoming electron:
\be
\begin{split}
	S^{(0)}&=-i\int\ud^4 x\ e^{i(p'+a_\infty).x}\bar{u}_{p'}(i{\fsl{D}}-m)\Psi_p^{\text{in}}(x) \\
	&=(2\pi)^3\delta^3_{\LCperp,\LCm}(p'+a_\infty-p)\bar{u}_{p'}\frac{\fsl{k}}{2k_\LCp}u_p \times \\
	&\times i\int\limits_{-\infty}^f\!\ud\phi\ (V_\infty-V)\exp\bigg(i\int\limits_0^{\phi}V_\infty-V\bigg) \;.
\end{split}
\ee
The integral over $\phi$ needs to be regulated. Inserting a small convergence factor we find, for $a\not=0$,
\be\label{SWE}
	S^{(0)}=(2\pi)^3\delta^3_{\LCperp,\LCm}(p'+a_\infty-p)\bar{u}_{p'}\frac{\fsl{k}}{2k_\LCp}u_pe^{i\theta} \;.
\ee
where
\be\label{THETA}
\theta:=\int\limits_0^{\phi_f}(V_\infty-V) \;.
\ee
We now apply (\ref{absolutSISTA}). The spin sum and average in our case gives
\be
\frac{1}{2}\cdot \frac{1}{4k_\LCp^2}\mathrm{Tr}[(\fsl{p}+m)\fsl{k}(\fsl{\pi}+m)\fsl{k}]=4p_\LCm^2 \;,
\ee
and it follows that the total probability is
\be\begin{split}
	\mathbb{P} &=\int\!\frac{\ud^3{{\bf p}'}}{(2\pi)^3 2p'_0} \frac{1}{2p_\LCm} \cdot (2\pi)^3\delta^3_{\LCperp,\LCm}({\sf p}' -{\sf p}) \cdot 4 p_\LCm^2 \\
	&=\int\!\frac{\ud^3{\sf p}'\theta(p_\LCm)}{2p'_-} 2p_\LCm\delta^3_{\LCperp,\LCm}({\sf p}' -{\sf p}) = 1\;,
\end{split}
\ee
where we used the Lorentz invariance of the measure in the second line to change variables. The reason why the probability is unity is discussed in Appendix \ref{APP-IR}; it is a manifestation of the IR problem.
% 
%%%%%%%%%%%%%%%%%%%%%
%%%%%%%%%%%%%%%%%%%%%
\section{IR structure}\label{APP-IR}
%%%%%%%%%%%%%%%%%%%%%
%%%%%%%%%%%%%%%%%%%%%
%
Here we establish in general how soft photons affect a given Feynman diagram in the Furry picture. We focus on the leading order IR divergences, following Weinberg's treatment \cite{Weinberg}.

We use dimensional regularisation to take care of the IR divergences, working in $1+d$ dimensions with $d>3$. Noting that our background field singles out a lightlike direction $\phi = k.x\sim x^\LCp$, we follow \cite{Casher:1976ae} and place the extra dimensions into the $d-1$ directions transverse to the background. This means in particular that all the structure of the plane wave is preserved by the regularisation. The measures in position and momentum space are then, in lightfront coordinates,
\be\begin{split}
	\ud x &:=\frac{1}{2}\ud x^{\LCp}\ud x^{\LCm}\ud^{d-1}x^\LCperp \equiv \frac{\ud\phi_x}{2k_{\LCp}} \ud x^{\LCm}\ud^{d-1}x^\LCperp \;,\\
	\ud q &:=\frac{\ud^{d+1}q}{(2\pi)^{d+1}}=\frac{\ud q_\LCp}{\pi}\frac{\ud q_\LCm \ud q^{d-1}_\LCperp}{(2\pi)^d} \;.
\end{split}
\ee
As we are only interested in the IR, we assume there is a cutoff in place to take care of the UV divergences, regarding which we note the following. By employing a proper time representation, all propagators can be expressed in terms of heat-kernels, and these are easily continued to $d\not=3$ following \cite{Luscher:1982wf}. The short-time expansion of the heat-kernel then gives a very convenient method for identifying UV divergences even when background fields are present. See \cite{Kibble:1975vz} for an example.
\begin{widetext}

\subsection{Soft photon correction to an external line}
We concentrate on incoming electron lines, other external lines can be treated similarly. A Feynman diagram in which the incoming electron emits $n$ {\it soft} photons (which may be real or virtual, we consider both cases below) with small momenta $l_j$ contains the term
\be
\begin{split}
	E_n(x_{n+1}):=&(-ie)^n\int\ud x_n...\ud x_1\ G(x_{n+1},x_n)\gamma^{\mu_n}e^{il_n.x_n} \ldots G(x_2,x_1)\gamma^{\mu_1}e^{il_1.x_1}\Psi_p^{\text{in}}(x_1) \;.
\end{split}
\ee
We will first factor out the infrared divergent part. To begin, consider $n=1$,
\be
\begin{split}
	E_1(x_2)=e\int\!\ud x_1\!\int\!\ud q\ \K_{q2}\frac{\fsl{q}+m}{q^2-m^2+i\epsilon}\bar{\K}_{q1}\fsl{\epsilon}\K_{p1}u_p \exp\bigg(-iq.x_2+i(q+l_1-p).x_1-i\int\limits_1^2V_q-i\int\limits_0^1V_p\bigg) \;.
\end{split}
\ee
The $x^\LCm$ and $x^\LCperp$-integrals sets ${\sf q}={\sf p} -{\sf l}_1$. (In the absence of the field, we would also be able to perform the $x^\LCp$ integral to set $q^\mu=p^\mu-l^\mu$.) Since we are only interested in the soft sector, and in particular the divergent terms, we employ the usual eikonal approximation, replacing $k.q\to k.p$ in $\K$. Define now a new variable $t$ by $q^2-m^2=2k.qt\approx 2k.pt$.  Changing integration variable from $q_{\LCp}$ to $t$ we find, again to lowest order in $l$
\be
\begin{split}
	E_1(x_2)=\frac{e}{2k.p}\int\!\ud\phi_1\int\!\frac{\ud t}{2\pi}\K_{p2}\bigg[\frac{\fsl{p}+m}{t+i\epsilon}+\fsl{k}\bigg]\bar{\K}_{p1}\gamma^{\mu_1}\K_{p1}u_p \exp\bigg(-it(\phi_2-\phi_1)-i(p-l_1).x_2-i\int\limits_0^2V_p-i\int\limits_1^2\frac{l_1.\pi}{k.p}\bigg) \;.
\end{split}
\ee
We have condensed our notation further: when $\pi$ appears under a $\phi$ integral it means $\pi(p;\phi)$, and when it appears outside such integrals it means $\pi(p;\infty)$. Consider the term in square brackets. Performing the $t$ integral, the $\fsl{k}$-term leads to a delta function setting $\phi_2=\phi_1$. This is the contribution from the lightfront zero mode \cite{Heinzl:2000ht}, which is interesting in itself but does not contribute to the IR divergence and so we drop it.  The remaining term in the square brackets gives $\theta(\phi_2-\phi_1)$, which restricts the $\phi_1$ integral. The spin term can be simplified to $(\fsl{p}+m)\bar{\K}\gamma^{\mu}\K u=2\pi^{\mu}u$, and we then find
\be
	E_1(x_2)=\Psi_p^{\text{in}}(x_2)\ e^{il_1.x_2}\bigg(\frac{-ie}{k.p}\bigg)\int\limits_{-\infty}^{\phi_2}\!\ud\phi_1\pi_1^{\mu_1}\exp\bigg(-i\int\limits_{\phi_1}^{\phi_2}\frac{l_1.\pi}{k.p}\bigg) \;.
\ee
The calculation is easily extended, so that we obtain
\be\label{EN}
\begin{split}
	E_{n}(x)=\Psi_p^{\text{in}}(x)\exp\bigg(i\sum\limits_{j=1}^n l_j.x\bigg) \left(-\frac{ie}{k.p}\right)^n\int\limits_{-\infty}^{x}\! \ud\phi_n\ldots \int\limits_{-\infty}^2\ud\phi_1\prod\limits_{j=1}^n\pi_j^{\mu_j}\exp\bigg(-i\int\limits_j^x\frac{l_j.\pi}{k.p}\bigg) \;.
\end{split}
\ee
The leading term is the incoming Volkov solution. The evaluation of the remaining integrals, which contains the IR divergence,  depends on whether a) the leg is attached to a `hard' vertex and is therefore part of a multi-particle scattering process or b) the leg continues to an outgoing line, in which case there are only soft photons in the process. We now consider these two cases, which are illustrated in Fig.~\ref{TVAAFALL}.
\subsection{Between a soft and a hard place}
\begin{figure}[t!]
\includegraphics[width=5cm]{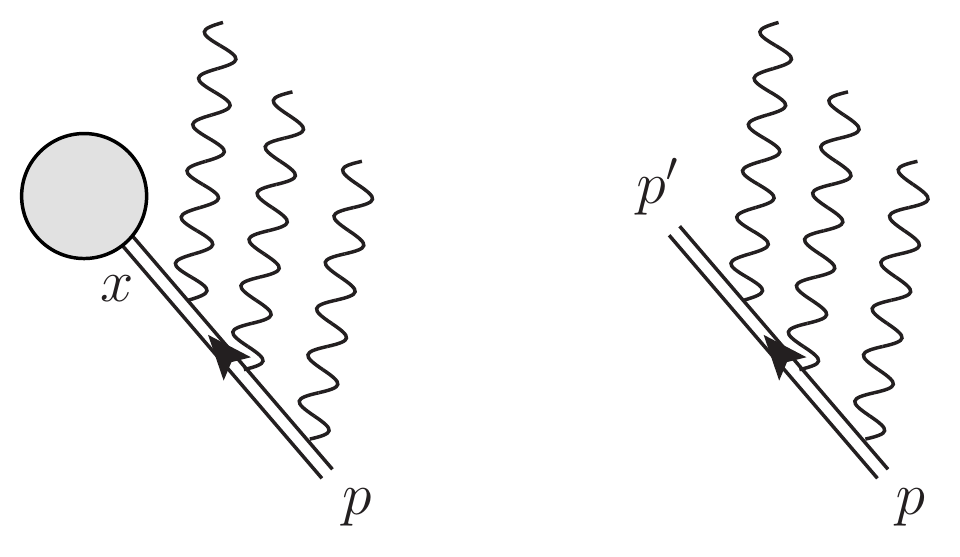}
\caption{\label{TVAAFALL} Left: an external leg emits {\it soft} photons (or emits and absorbs virtual photons) as part of a scattering process with a hard vertex at $x^\mu$. Right: emission of many {\it soft} photons from a single electron. There is no hard scattering part to this diagram.}
\end{figure}
We begin with the left hand diagram in Fig.~\ref{TVAAFALL}, considering the effect of the soft photon lines in (\ref{EN}) (which can correspond real emission or virtual loops, as we will need both) on a hard scattering process. We will see that in this case, the IR divergence does not depend on the properties of our background field.

So, we assume that $E_n(x)$ is connected to a hard vertex. In that case, there is a hard photon momentum in the exponent of (\ref{EN}) at position $x$, and relative to this we can neglect the term $\sum l_j$. Since we have reduced the spin terms to products of $\pi$'s in (\ref{EN}), the integrand therein is symmetric with respect to the $\phi_j$ except for the step functions. This simplifies when we sum over all the orders in which the soft photons can be emitted (i.e.\ when we account properly for all diagrams): 
\be
\begin{split}
&\sum\limits_{l-perm}E_n(x)=\Psi_p^{\text{in}}(x) \left(-\frac{ie}{k.p}\right)^n\prod\limits_{j=1}^n\int\limits_{-\infty}^x\ud\phi_j\pi_j^{\mu_j}\exp\left(-i\int\limits_j^x\frac{l_j.\pi}{k.p}\right) \;,
\end{split}
\ee
and we see that the step functions drop out. The divergent part of the $\phi$-integrals comes from the region before the pulse turns on. Since the leg is attached to the hard vertex, the upper limit of the integrals in the exponent is unimportant as long as it is finite, and we can make the following replacement without affecting the leading divergence,
\be
	\int\limits_{-\infty}^x\phi_j\rightarrow\int\limits_{-\infty}^0\phi_j \;.
\ee
Performing the $\phi$ integrals with the help of a convergence factor as usual, we find
\be
	\sum E_n=\Psi_p^{\text{in}}(x) \ \prod\limits_{j=1}^n\frac{-ep_j^{\mu_j}}{l_j.p-i\epsilon} \;,
\ee
in which the soft contributions factor off, meaning that the Feynman amplitude factorises into a hard part and a soft correction, as for QED without background fields. Similar expressions hold for other external lines. Hence, we have reduced the IR problem in plane waves to the case with no background, with no surprises thus far, and we may proceed as in \cite[\S 13]{Weinberg} to show that the leading infrared divergences cancel to all orders when one sums the probabilities for indistinguishable processes. These statements hold for processes in which a `hard part' can be identified, i.e.\ assuming that the external lines are connected to hard vertices. We turn now to single electron processes with only soft vertices.

\subsection{Soft processes}
The $S$-matrix element for a diagram with a single electron and $n$ soft vertices, see the right hand diagram of Fig.~\ref{TVAAFALL}, contains
\be
	S_n:=-ie\int\!\ud x\ \bar{\Psi}_{p'}^{\text{out}}(x)\gamma^{\mu}e^{il.x}E_{n-1}(x) \;.
\ee
Since the photons are soft, $k.q\approx k.p$ and $k.p'\approx k.p$. The spin terms can then be simplified and taken outside the integrals, leaving
\be
S_n=(2\pi)^d \delta^3_{\LCm,\LCperp}(p'+a_\infty-p)e^{i\theta}\bar{u}_{p'}\frac{\fsl{k}}{2k_{\LCp}}u_p\left(-\frac{ie}{k.p}\right)^n \int\limits_{-\infty}^{\infty}\!\ud\phi_n\int\limits_{-\infty}^n\!\ud\phi_{n-1}\ldots\int\limits_{-\infty}^2\!\ud\phi_1\prod\limits_{j=1}^n\pi_j^{\mu_j}\exp\bigg(i\int\limits_0^j\frac{l_j.\pi}{k.p}\bigg) \;,
\ee
with $\theta$ as in (\ref{THETA}). (We drop the factor $\sum l_j$ from the delta functions. In a more rigorous treatment the soft photon energies should be restricted so that this sum is less then some given energy, see \cite{Weinberg}. The divergent part is still the same.) As before, $S_n$ simplifies when we sum over the permutations of the $l_j$,
\be\label{SN}
	\sum\limits_{l-perm}\! S_n=(2\pi)^d\delta^3_{\LCm,\LCperp}(p'+a_\infty-p)e^{i\theta}\bar{u}'\frac{\fsl{k}}{2k_{\LCp}}u \left(-\frac{ie}{kp}\right)^n\prod_{j=1}^n\int\ud\phi_j\pi_j\exp\left(i\int\limits_0^j\frac{l_j.\pi}{k.p}\right) \;.
\ee  
We have derived this formula for $n\geq 1$ (with $E_0=\Psi^{\text{in}}$), but it also holds for $n=0$, when it describes elastic scattering at tree level, see Appendix \ref{APP1} above. The significant difference compared to the case of hard-soft factorisation is that in soft processes, the outgoing electron's momentum is fixed by classical momentum conservation, in other words by the properties of the background field, and in particular $a_\infty$.

The soft photons can be real or virtual. For each real emission we multiply (\ref{SN}) by a polarisation vector $\epsilon$, giving
\be
	-\frac{ie}{k.p}\int\ud\phi\ \epsilon.\pi\exp\left(i\int\limits^{\phi}\frac{l.\pi}{k.p}\right)=e\epsilon\left(\frac{\pi}{l.\pi}-\frac{p}{l.p}\right) \;.
\ee
\end{widetext}
At the level of the probability we sum over polarisations, which gives minus the above expression squared, and then integrate over the photon momenta. We get the same factor for each photon, but with a symmetry factor of $1/n!$. The contribution from the emission of $n$ soft photons is therefore
\be
\begin{split}
	\frac{1}{n!}\bigg[-e^2\int\frac{\ud^dl}{(2\pi)^d}\frac{1}{2l_0}\left(\frac{\pi}{l.\pi}-\frac{p}{l.p}\right)^2\bigg]^n \;,
\end{split}
\ee
For each virtual photon we choose $l_j=-l_i=l$ and multiply (\ref{SN}) by
\be
	\int\ud l\frac{-ig_{\mu_i\mu_j}}{l^2+i\epsilon} \;.
\ee
We then have
\be\label{VP}
	\int\ud l\frac{-i}{l^2+i\epsilon}\left(-\frac{ie}{k.p}\right)^2\int\!\ud\phi_i\ud\phi_j\pi_i\pi_j\exp\bigg(-i\int\limits_j^i\frac{l.\pi}{k.p}\bigg) \;.
\ee
We divide this into two parts, $\phi_i>\phi_j$ and $\phi_i<\phi_j$, and change variable $l\rightarrow -l$ in the $<$-part. We can then close the $l_0$-contour in the lower half plane, and (\ref{VP}) becomes
\be\label{B17}
\int\frac{\ud^dl}{(2\pi)^d}\frac{1}{2l_0}\left(\frac{e}{k.p}\right)^2\int\ud\phi_i\ud\phi_j\pi_i\pi_j(\theta_>e^-+\theta_<e^+) \;,
\ee
with obvious notation. The imaginary part of (\ref{VP}) diverges like  
\be
	\text{Im} \sim\int\ud\phi \;,
\ee
but we will shortly see that it drops out of probabilities. In the real part, the $\phi_i$ integrals can be performed and are finite, and (\ref{VP}) becomes
\be
e^2\int\frac{\ud^dl}{(2\pi)^d}\frac{1}{2l_0}\left(\frac{\pi}{l.\pi}-\frac{p}{l.p}\right)^2+i...
\ee
We get one such factor for each virtual photon with a factor of $1/2^n n!$, which is the number of identical permutations of the sum over $l_j$ for $n$ virtual photons. Summing over all $n$ we get the all-orders loop contribution to a soft process,
\be\label{SV}
\begin{split}
\exp\left(\frac{e^2}{2}\int\frac{\ud^dl}{(2\pi)^d}\frac{1}{2l_0}\left(\frac{\pi}{l.\pi}-\frac{p}{l.p}\right)^2+i...\right) \;.
\end{split}
\ee
This contribution appears mod-squared at the level of the probability, which removes both the leading factor of one half, and the divergent imaginary term; the latter are the usual phase divergences. Hence, returning to the example of Appendix~\ref{APP1}, the probability for elastic scattering including all soft loop contributions is given by the modulus squared of (\ref{SV}), 
\be\label{X-derive}
	\mathbb{P}=\exp\bigg[e^2\int\frac{\ud^dl}{(2\pi)^d}\frac{1}{2l_0}\bigg(\frac{\pi}{l.\pi}-\frac{p}{l.p}\bigg)^2\bigg] \;.
\ee
When $a_\infty\not=0$,  i.e.\ when the background field is unipolar, the loops give an IR divergent contribution as $d\to 3$.

\end{document}